# Engineering the geometry of stripe-patterned surfaces towards efficient wettability switching

**Michail E. Kavousanakis[1,a], Carlos E. Colosqui[2,b] and Athanasios G. Papathanasiou[*1,c]**

[1]School of Chemical Engineering, National Technical University of Athens, 15780, Greece.
[2] Benjamin Levich Institute of Physico-Chemical Hydrodynamics, City College of New York, New York, NY 10031, USA
[a] E-mail: mihkavus@chemeng.ntua.gr.  [b] E-mail: ccolosqui@ccny.cuny.edu.
[*c] Corresponding author. E-mail: pathan@chemeng.ntua.gr; Web: http://www.chemeng.ntua.gr/people/pathan.
Fax: +30 210772 3298; Tel: +30 210 772 3234

**Abstract**

The ability to control wettability is important for a wide range of technological applications in which precise microfluidic handling is required. It is known that predesigned roughness at a micro- or nano- scale enhances the wetting properties of solid materials giving rise to super-hydrophobic or super-hydrophilic behavior. In this work, we study the dependence of the apparent wettability of a stripe-patterned solid surface on the stripe geometry, utilizing systems level analysis and mesoscopic Lattice-Boltzmann (LB) simulations. Through the computation of both stable and unstable states we are able to determine the energy barriers separating distinct metastable wetting states that correspond to the well-known Cassie and Wenzel states. This way the energy cost for inducing certain wetting transitions is computed and its dependence on geometric features of the surface pattern is explored.

**Keywords**

Superhydrophobic, energy barriers, bifurcation analysis, Lattice-Boltzmann simulation

## 1. Introduction

The design of solid surfaces with fully tunable wetting properties is of critical importance for many technological applications including: biomimicking materials with super-hydrophobic or superhydrophilic properties [1, 2], microfluidic devices [3, 4], optics (liquid lenses) [5] and low consumption displays [6]. Wetting modification can be achieved with the use of smart coating materials, which are responsive to irradiation, pH variation, thermal and electric actuation [7, 8]. However, certain limitations originate from the chemistry of the material [9] (e.g., fluoropolymer top coatings exhibit water contact angles up to 120$^o$). Surface roughness can enhance the wetting properties of smooth solids [10]; e.g., artificially constructed micro- or nano-scale surface structures can exhibit contact angles close to the upper limit of non-wetting by water droplets when coated with fluoropolymers [11].

These surfaces are commonly pillared or honeycomb cells structures. Superhydrophobicity is achieved when the droplet sits on top of the protrusions attaining a nearly spherical shape; this state is commonly referred to as 'Cassie state' [10] and the droplets are characterized 'Fakir droplets', with apparent contact angle values exceeding 150$^o$; this 'living on air' state is reminiscent of a fakir sitting on a bed of nails (surface protrusions) [12]. Impaled states, commonly called 'Wenzel' states [10], can also be admitted on rough solid surfaces, in response to the demand for materials with strong affinity to water [13]. A Wenzel state is –usually-energetically more favorable than a Cassie state, for which appropriate perturbations cause the droplet to impale the surface protrusions forming sticky states of hydrophilic wetting behavior.

Cassie and Wenzel states correspond to distinct minima of the free energy, and depending on the material wettability, as well as the surface roughness geometry, one or the other state is energetically more favorable [14, 15]. The physical picture of having two thermodynamically (meta)stable states (Wenzel and Cassie) suggests the existence of an intermediate unstable state, which corresponds to a saddle of the free energy landscape with a one-dimensional unstable manifold, where mechanical equilibrium is achieved. The free energy difference between this saddle and the local minima corresponding to (meta)stable equilibrium states determines the energy barrier, that a perturbation must surpass in order to induce a transition between wetting states. Furthermore, the surface roughness morphology has a significant effect on the magnitude of energy barriers separating the different wetting states [16] (e.g. mushroom-like pillars enhance the robustness of superhydrophobic behavior [17]).

Experimental evidence shows that transition between Cassie and Wenzel states is in practice one-way process [18, 19]. Material wettability enhancement of a rough surface, that sustains a Cassie state, can effectively induce collapse transitions (e.g., through the electrowetting effect [3]). However the reverse transition usually requires strong external actuation as reported in Krupenkin et al. [20], where extreme heating is applied. Such a strong thermal actuation is prohibitive for handling sensitive liquids as required in biofluidics.

Thus, optimal switching between wetting states is a matter of suitable surface roughness design, that minimizes the corresponding energy barriers, and suitable actuation that delivers the required energy without significant losses. On the contrary, surfaces that maintain a robust hydrophobic or hydrophilic behavior, requires roughness design that maximizes the energy barriers. Hence, the design of surfaces boils down to an energy barrier minimization / maximization problem, in which the optimization parameters are the geometric characteristics of the surface roughness.

For this purpose, we have recently developed a time-stepper based technique [21] that enables the tracking of entire branches of steady-state solutions, utilizing lattice Boltzmann (LB) simulations [22, 23]. By adopting the computational framework presented in [24, 25] we perform systems-level numerical tasks, e.g., bifurcation and stability analysis, which enables the computation of all admissible wetting states on a model structured surface. Energy barrier computations are performed by accounting for the entire droplet shape, i.e. not for a unit cell [17]. In addition, no assumptions are made regarding the pinning of the droplet contact line at each surface protrusion edge.

The article is organized as follows: we first provide a brief description of the mesoscopic LB model, utilized to simulate wetting phenomena on micro-structured solid surfaces. The computational framework, which enables the implementation of systems-level analysis using dynamic (fine scale) simulators, is described in Section 2.2. The theoretical background for energy barrier computations is described in Section 2.3. In the Results Section, we show the efficiency of our proposed computational methodology to track entire solution branches, which can be admitted on corrugated solid surfaces. We study the effect of various geometric features of stripe-patterned surfaces, such as the stripe edge sharpness, height and width, as well as the distance between stripes, on the apparent wetting behavior. Finally, we investigate the effect of these geometric features on facilitating or impeding wetting transitions through the computation of energy barriers. In Section 4, we summarize our main findings, and outline potential future directions.

## 2. Methods

We model the wetting of micro-structured solid surfaces using dynamic simulations with the LB method. LB simulation is advantageous over alternative computational methods when dealing with solid surfaces of geometric complexity. Our model [22, 26] incorporates fluid-fluid and solid-fluid interactions that give rise to surface tension, disjoining pressure, and capillarity.

*2.1 Mesoscopic simulation of wetting phenomena in solid surfaces*

In the simulations reported here, we solve the Boltzmann-BGK equation for a single component fluid:

$$\frac{\partial f}{\partial t} + \mathbf{v} \cdot \nabla f = -\frac{f - f^{eq}}{\tau} + \frac{\delta f}{\delta t}, \qquad (1)$$

where $f(\mathbf{x},\mathbf{v},t)$ is the single-particle distribution function defined in space $(\mathbf{x},\mathbf{v})$ and $\tau$ is the relaxation time determining the kinematic viscosity of the fluid. The equilibrium distribution:

$$f^{eq}(\mathbf{x},\mathbf{v},t) = \frac{\rho}{(2\pi\theta)^{D/2}} \exp\left[-(\mathbf{v}-\mathbf{u})^2/(2\theta)\right], \qquad (2)$$

is given by a Maxwell-Boltzmann distribution, where $\theta = k_B T/m$ is the specific thermal energy (*T*: temperature, $k_B$: the Boltzmann constant, *m*: molecular mass) and $\mathbf{u}$ is the fluid velocity defined in configuration space $(\mathbf{x},t)$. In the absence of external fields, the forcing term $\delta f/\delta t$ in Eq. (1) is a functional of pseudo-potentials [27] for fluid-fluid and fluid-solid interactions [22].

The interactions modeled by the fluid-fluid pseudo-potential employed here [26] produce separation in two ideal phases, i.e., the (dense) liquid and (light) ambient phase, and control the magnitude of the fluid-fluid interfacial tensions [28, 29]. Thus, the static contact angle (Young's angle), $\theta_Y$, is tuned by the attraction and repulsion parameters of the fluid-solid pseudo-potential. A detailed description of the LB model and pseudo-potentials employed is provided in Colosqui et al. [22].

*2.2 Systems level analysis utilizing LB simulations*

Lattice Boltzmann simulations can provide an accurate dynamic description of wetting on micro-structured surfaces. However, LB simulations can only converge to steady states ((meta)stable wetting states) that are dynamically stable, lacking the ability to compute steady states that are dynamically unstable. Such unstable states determine the energy barriers separating distinct wetting states. Moreover, computing all steady state solutions for multiple material wettabilities is computationally intensive through direct dynamic simulation.

Such limitations have been overcome through a time-stepper based computational framework, described in reference [21], which wraps around the LB dynamic simulator and enables the computation of steady state solutions, utilizing relatively short and appropriately initialized LB executions. Employing a state vector $\mathbf{U}$, constructed with coarse-grained variables (e.g., mass density, momentum and stresses) reported by the LB simulations at discrete time instances, one can construct a discrete time map

$$\mathbf{U}(t+T) = \Phi_T(\mathbf{U}(t);\lambda), \qquad (3)$$

where $T$ is the reporting time horizon and $\lambda$ is a continuation parameter, which in our study is the Young's contact angle (i.e., $\lambda \equiv \theta_Y$). In this work, **U** is constructed with the leading order moments of the single-particle distribution function, $f$, namely the mass density, $\mathbf{M}^{(0)}(\mathbf{x},t) = \int f(\mathbf{x},\mathbf{v},t)d\mathbf{v} = \rho$, the fluid momentum, $\mathbf{M}^{(1)}(\mathbf{x},t) = \int f(\mathbf{x},\mathbf{v},t)\mathbf{v}d\mathbf{v} = \rho\mathbf{u}$, and the momentum flux tensor, $\mathbf{M}^{(2)}(\mathbf{x},t) = \int f(\mathbf{x},\mathbf{v},t)\mathbf{v}\mathbf{v}d\mathbf{v} = \rho\mathbf{u}\mathbf{u} + \Pi$, ($\Pi$: stress tensor).

For isothermal conditions, the distribution $f$ can be approximated as a function of its moments $\mathbf{M}^{(0)}$, $\mathbf{M}^{(1)}$, and $\mathbf{M}^{(2)}$ (for details see references [21, 22, 30]). This approximation entails that no information is lost when initializing $f$, the mesoscopic variable for LB simulation, from information contained in the state vector **U**, the coarse-grained variable evolved by the time stepper. Since we have ensured the lossless transfer of information between the different levels of description, one can effectively perform systems level analysis. In particular, we can compute the steady state solutions $\mathbf{U}^*$ by solving the following non-linear system:

$$\mathbf{R} \equiv \mathbf{U}^* - \Phi_T(\mathbf{U}^*;\lambda) = \mathbf{0}. \tag{4}$$

Since the $\Phi_T$ operator is not readily available, time-stepper based techniques are adopted [24, 31-33] for the solution of the non-linear system of equations (4) (e.g., the matrix-free Newton-GMRES method [32]).

Extensions of this application enable to perform bifurcation analysis by means of pseudo-arc-length parameter continuation techniques [34]. Such techniques enable the computation of all stable and unstable steady states, as well as stability analysis using matrix-free eigensolvers (Arnoldi eigensolvers [35]), which compute the leading eigenvalues of the Jacobian matrix, $\partial\Phi_T(\mathbf{U}^*)/\partial\mathbf{U}^*$. When at least one eigenvalue crosses the unit circle in the complex plane, then the solution is characterized as dynamically unstable [36] since even a small perturbation applied on the steady state solution $\mathbf{U}^*$ will eventually cause the transition to a different (stable) steady state solution.

*2.3 Energy barrier computation*

Stable steady states correspond to distinct minima of the surface free energy. When multiple stable steady states co-exist (for the same parameter value $\lambda \equiv \theta_Y$), at least one intermediate unstable state also exists. This unstable state (undetectable in experiments) corresponds to a saddle of the surface free energy; the difference between the saddle and the local energy minima determines the energy barrier for the realization of wetting (or de-wetting) transitions.

For small liquid droplets (<10 μL), the effect of gravity can be neglected, thus the free energy can be expressed as

$$F = \gamma_{LV} A_{LV} + \gamma_{LS} A_{LS} + \gamma_{SV} A_{SV}, \tag{5}$$

where $\gamma_{LV}$, $\gamma_{LS}$ and $\gamma_{SV}$ are the interfacial tension of the Liquid-Vapor (LV), Liquid-Solid (LS) and Solid-Vapor (SV) interfaces, respectively; $A_{LV}$, $A_{LS}$ and $A_{SV}$ are the surface areas of the LV, LS and SV interfaces (see Figure 1).

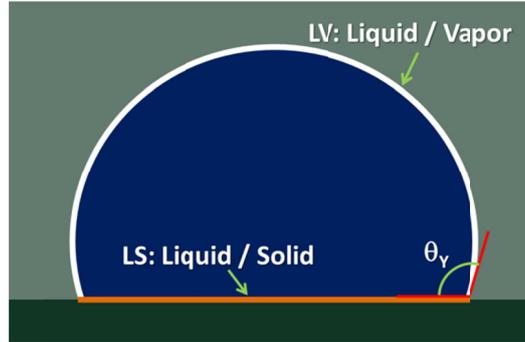

**Figure.1** Schematic representation of Liquid/Vapor (LV) and Liquid/Solid (LS) interfaces of a cylindrical droplet wetting a flat solid surface with Young's contact angle, $\theta_Y$.

If we denote with $F_s$, $F_c$, and $F_u$ the surface free energy of the stable suspended (s), stable collapsed (c), and unstable (u) – intermediate wetting steady states, which can be admitted on a structured solid surface with Young's contact angle $\theta_Y$ (see Figure 2) then a wetting (or suspended to collapsed state, s→c) transition can be computed from:

$$E_{s \to c} = F_u - F_s \to \frac{E_{s \to c}}{\gamma_{LV}} = (A_{LV,u} - A_{LV,s}) - \cos\theta_Y (A_{LS,u} - A_{LS,s}), \tag{6}$$

by taking into account Young's equation, $\gamma_{LV} \cos\theta_Y + \gamma_{LS} = \gamma_{SV}$. Thus, the energy barrier computation requires only the surface area of the LV and LS interfaces of the unstable and stable wetting steady states for a given material wettability (i.e., Young's contact angle value $\theta_Y$). The energy barrier for a de-wetting transition, $E_{c \to s}$, is computed similarly.

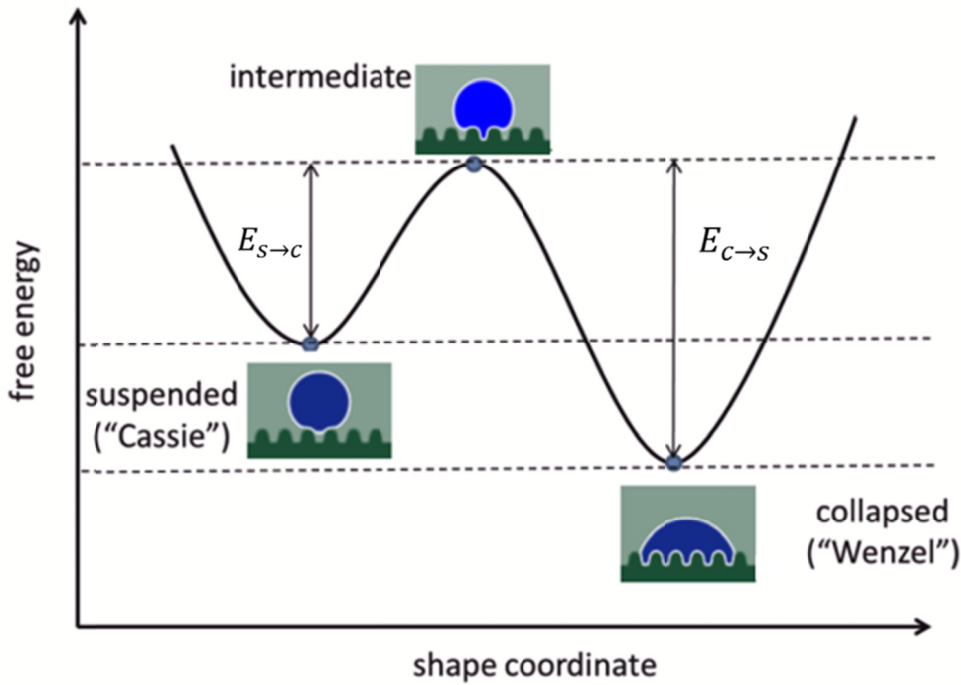

**Figure 2.** Schematic of free energy landscape for a corrugated solid surface and constant Young's contact angle, $\theta_Y$.

## 3. Results and discussion

### 3.1 Bifurcation analysis

We investigate the effect of structure geometry on the wetting properties of corrugated solid surfaces. In particular, we perform parametric continuation analysis, with $\theta_Y$ serving as the continuation parameter, to compute all possible wetting states that can be admitted on six-striped solid surfaces (see Fig. 3). Under simulated conditions, we set a density ratio of 10:1, and a compressibility ratio of 15.6:1 in thermodynamic equilibrium [22]. As reported above, the Young's contact angle value is a function of the attraction and repulsion parameters in the interaction pseudo-potentials that model fluid-solid interactions.

The simulated problem is two dimensional, and the corrugated surface can be considered as extending infinitely along the direction normal to the *xz*-plane, as illustrated in Fig. 3. The shape of the employed $N(=6)$-striped surface is determined from:

$$y = \frac{h}{2}\sum_{i=1}^{N/2} \tanh\left(\frac{r - r_{start}}{p}\right) - \tanh\left(\frac{r - r_{end}}{p}\right), \tag{7}$$

where $r_{start}=d/2+(i-1)(w+d)$ and $r_{end}=x_{start}+w$; $h$ determines the height of each individual protrusion, and $p$ is the sharpness factor (decreasing $p$, increases the stripe edge sharpness); $w$ is the width of each stripe's basis and $d$, determines the distance between stripes; $r$ denotes the Euclidean distance from the center, $x_o$, of the *x*-domain: $r=|x-x_o|/2$. The solid structures under study are of comparable size with the droplet diameter. In the results presented below, $h$, $w$, and $d$ are reported in dimensionless units by normalizing with the droplet radius.

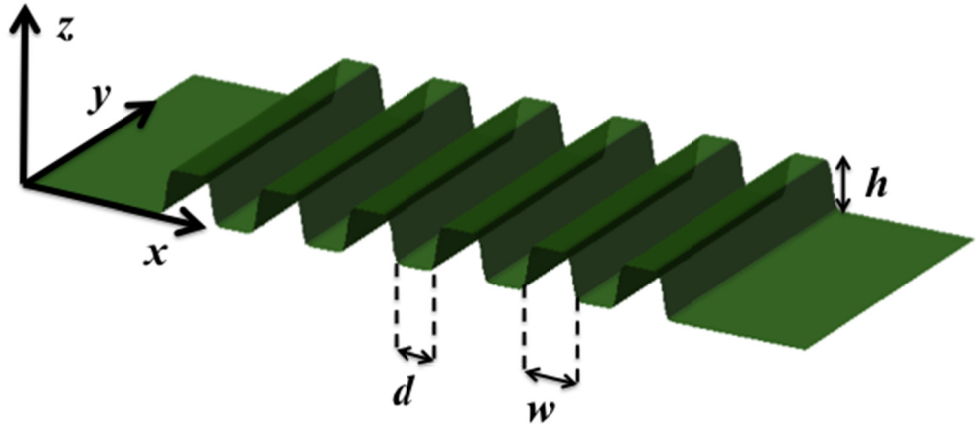

**Figure 3.** Schematic representation of a six-striped solid surface extending infinitely along the *y*-direction.

In Kavousanakis et al. [21], we demonstrated how one can systematically explore the solution space of wetting states that can be admitted on structured surfaces, utilizing LB simulations. Through the application of arc-length parameter continuation techniques [34] we performed bifurcation analysis and traced entire equilibrium solution families, checking also their dynamic stability.

A representative bifurcation diagram, which shows the wetting behavior as a function of Young's contact angle, $\theta_Y$, on a six-striped solid surface is depicted in Fig. 4. One can observe the existence of an interval of $\theta_Y$ values, within which five distinct wetting steady states can co-exist, three of them being stable (solid lines in Fig. 4), and two of them being unstable steady states (dashed lines in Fig. 4).

Hydrophobic materials with $\theta_Y > 110º$ give rise to superhydrophobic-'Cassie' states, where the droplet is suspended on two stripes (branch (I)). As we gradually increase wettability of the solid material, a collapse transition leading to states of droplets wetting four stripes occurs at $\theta_Y \approx 76°$ (point A), while a second collapse transition from branch (III) to branch (V), which corresponds to states of droplets wetting all six stripes, requires further lowering of $\theta_Y$ down to 20° (point B). Branch (III) corresponds to intermediate stable steady state solutions, commonly known as Cassie or Wenzel impregnating states [37, 38]; such states exhibit "sticky" properties and are characterized by high contact angle hysteresis, and appears in the bifurcation diagram in the form of a long range of $\theta_Y$ values over which bistability is observed.

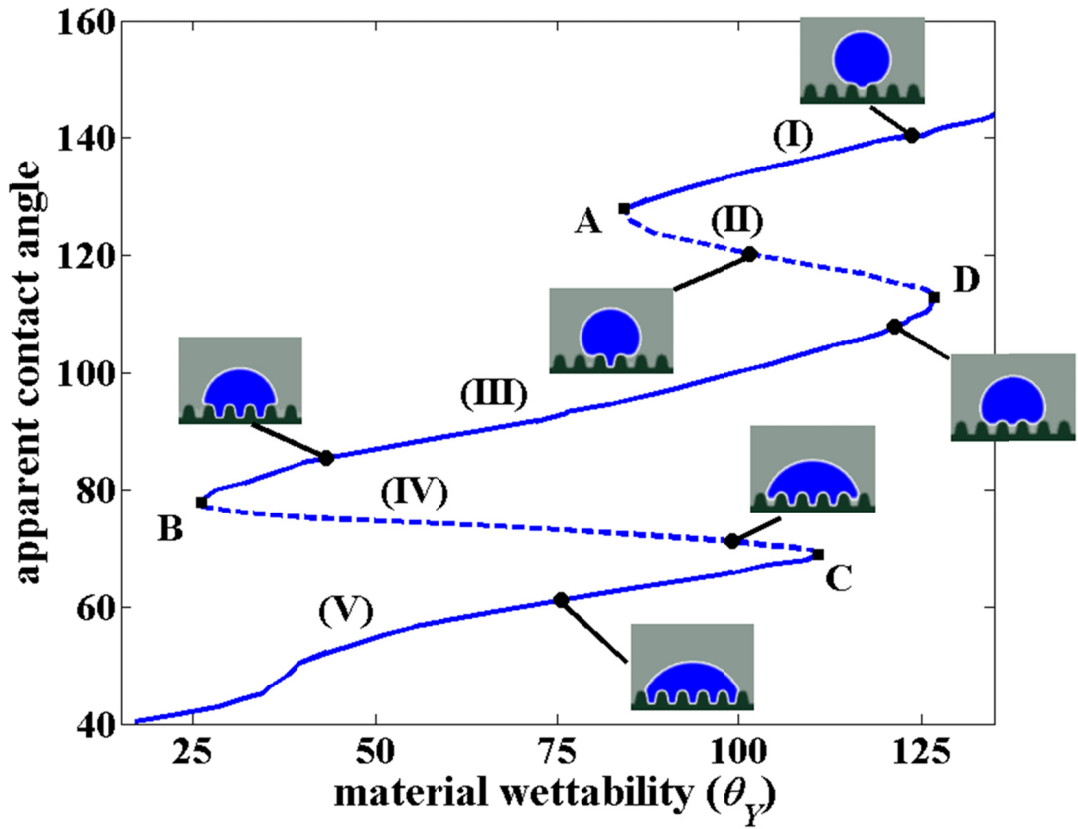

**Figure 4.** Bifurcation diagram showing the dependence of wetting steady states on a six-striped solid surface as a function of $\theta_Y$. Width $w=0.375$, distance $d=0.375$, height $h=0.5$, and sharpness factor $p=2$.

In a reverse experiment, starting from hydrophilic states (branch (V)), and gradually decreasing the solid material wettability, a first de-wetting transition towards states of a droplet wetting four stripes would require an increase of $\theta_Y$ at 111° (point C in Fig. 4), and a further increase to $\theta_Y = 124°$ (point D) for a (III) to (I) de-wetting transition. Hence, the transition from suspended to collapsed states is hysteretic and the range of hysteresis is determined by the length of the branch of unstable solutions (branches (II) and (IV)).

In this work, we investigate the effect of various geometric features of the structured solid surface on its apparent wetting properties. Unless stated otherwise $w=d=h=0.375$, and the sharpness factor is $p=2$.

*3.1.1 Sharpness effect*

The effect of stripe edge sharpness is shown in Fig. 5. The parameter $p$ in Eq. (7) controls the curvature of lateral walls (sharpness). A sharpness factor $p=1$ corresponds to sharp edges while $p=2$ corresponds to smoother stripe edges. Sharpness has a negligible effect on the first "(I) to (III)" wetting transition turning point ($\theta_Y$ =84.5° for sharp and 84.2° for smooth stripes); a more prominent effect is observed in the "(III) to (V)" transition point. Sharp stripes require lowering of wettability down to $\theta_Y$ =16° as opposed to $\theta_Y$ =26° for smoother stripes. Branch (IV), for sharp stripes, extends over a range of 91.9°, which is significantly higher compared to the corresponding range of 81.5° for smoother corrugations. This difference can be attributed to pinning of the contact line at the sharp edges of the stripes.

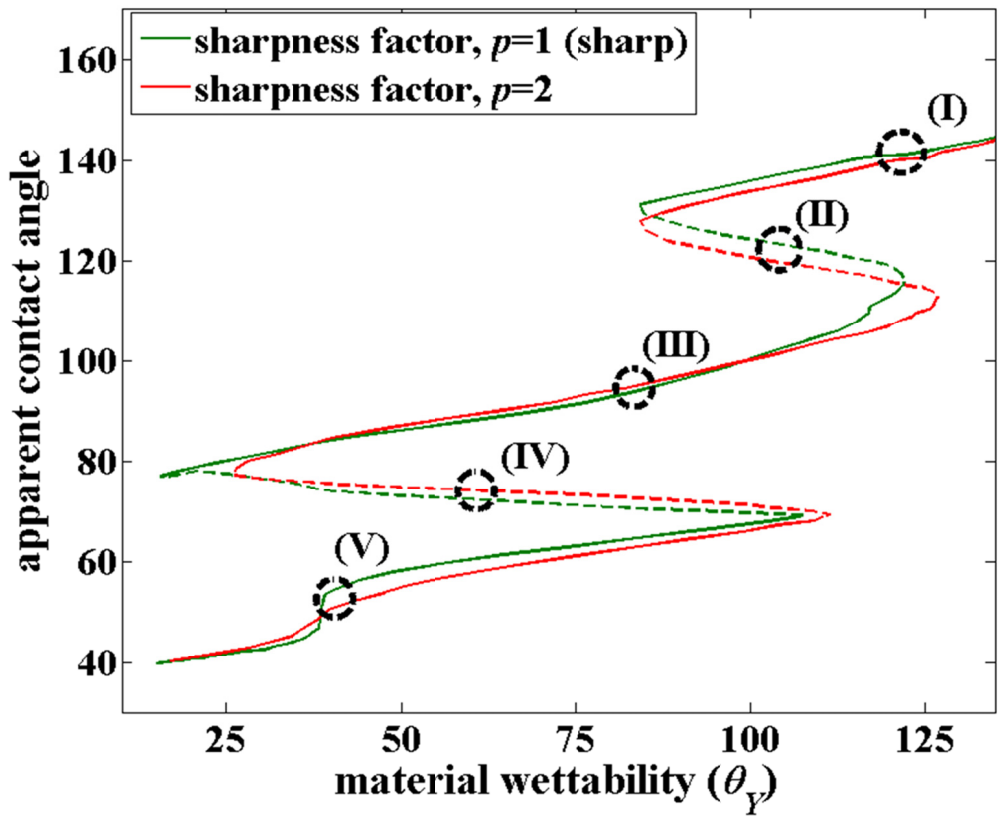

(a)

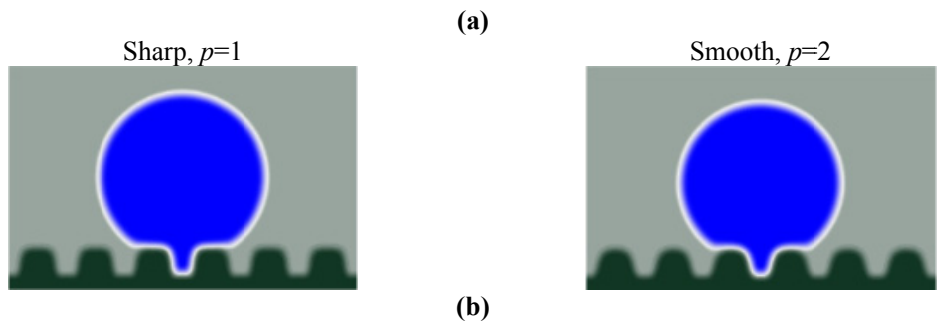

(b)

**Figure 5. (a)** Effect of stripe edge sharpness on the apparent wetting properties of a corrugated solid surface. **(b)** Two representative wetting steady state solutions corresponding to $\theta_Y$ =100° at branch (I). Width $w$=0.375, distance $d$=0.375, and height $h$=0.375.

*3.1.2 Distance effect*

In Fig. 6 we present the effect of varying the distance between the stripes on the wetting properties of a corrugated solid surface. Notably, the range of bistability increases significantly by increasing the strip distance. Besides a leftward shift of the wetting transition turning points when increasing distance, $d$, the de-wetting transition points are located at higher $\theta_Y$ values. In particular, the critical value of $\theta_Y$ for "(I) to (III)" and "(III) to (V)" wetting transitions are located at 60.8° and 15.5° respectively, for $d$=0.450, as opposed to 84.2° and 26° respectively for $d$=0.375. The critical values of $\theta_Y$ signaling de-wetting "(V) to (III)" and "(III) to (I)" transitions for $d$=0.450 are 112.7° and 128.6°, while the corresponding critical values for $d$=0.375 are 111.4° and 126.7°, respectively. Thus, the hysteresis range for transitions between the suspended states (branch (I)) and the collapsed states (branch (V)) gets enlarged when the distance between stripes is increased.

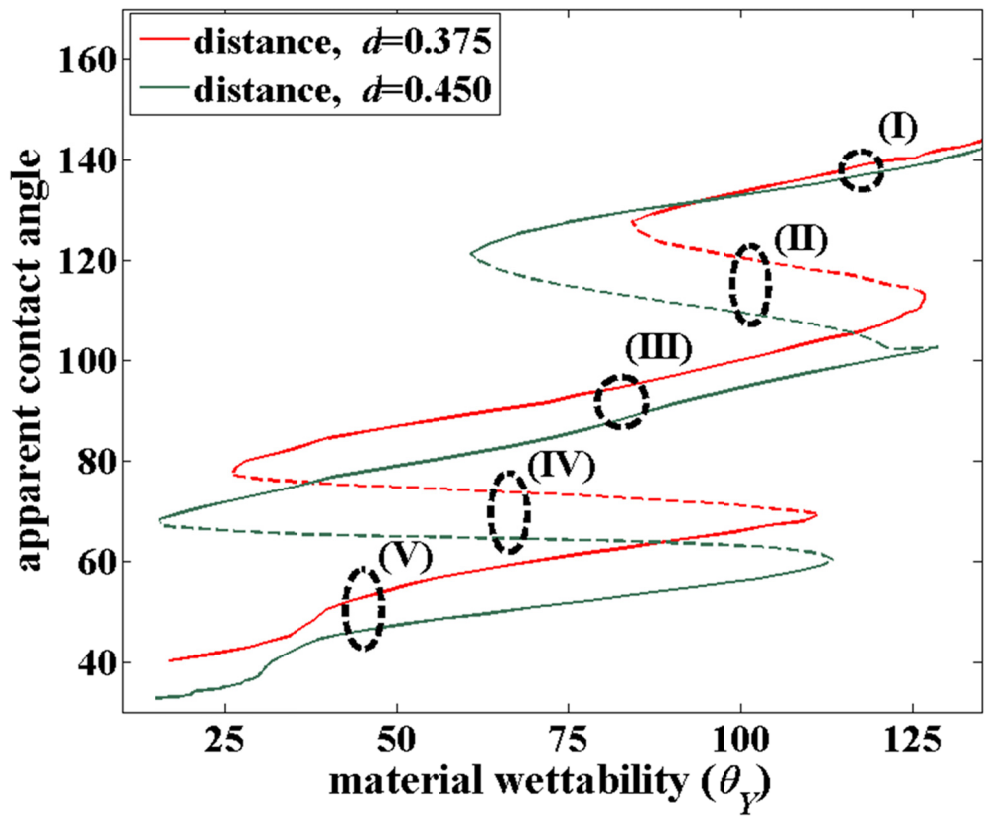

(a)

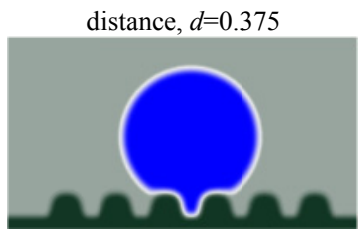
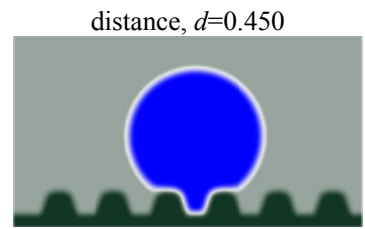

(b)

**Figure 6.** (a) Effect of varying the distance between stripes on the wetting properties of a corrugated solid surface. (b) Branch (I) wetting steady state solution at $\theta_Y$=100°. Width $w$=0.375, height $h$=0.375 and sharpness factor $p$=2.

*3.1.3 Height effect*

The effects of varying the stripe height are illustrated in the bifurcation diagram of Fig. 7. By increasing the depth of corrugations, the critical $\theta_Y$ contact angle values for wetting transitions decrease, thus suggesting that sufficiently deep corrugations can prevent wetting transitions. The critical $\theta_Y$ values for the reverse de-wetting transition remain practically unchanged when the height is increased from $h=0.375$ to $0.5$ (i.e., about a 33% increase).

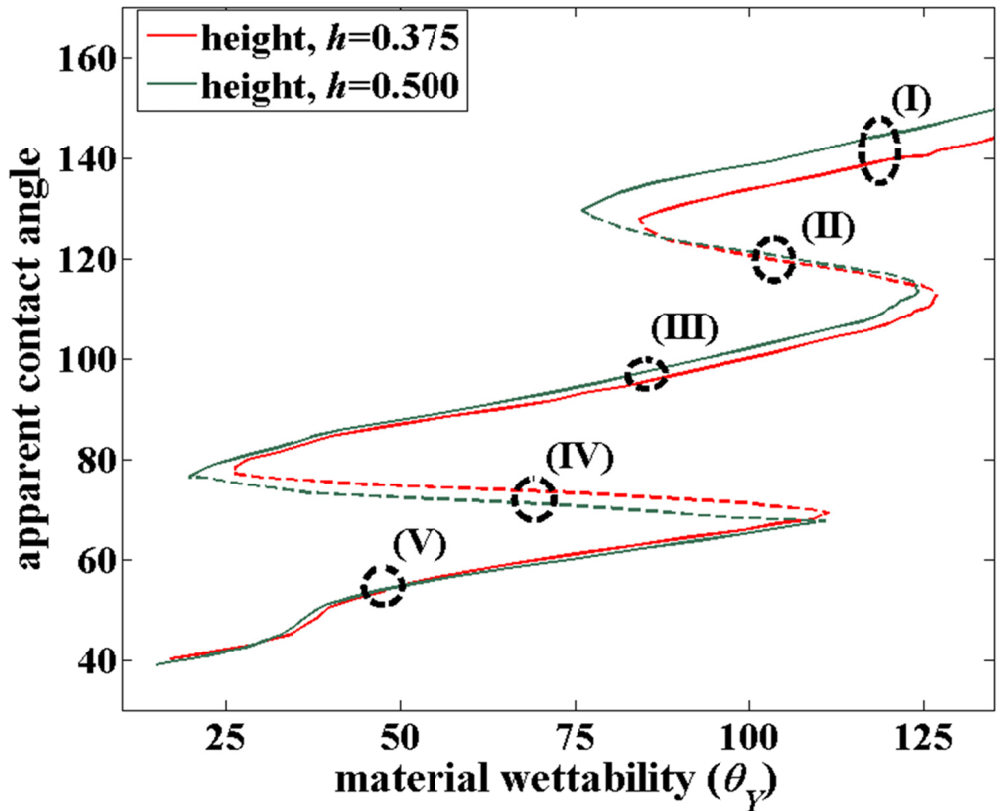

(a)

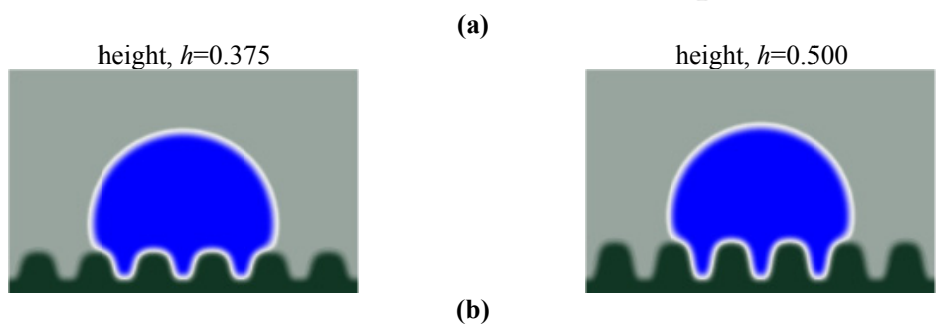

(b)

**Figure 7.** (a) Effect of height variation on the apparent wetting properties of a solid surface decorated with six stripes. (b) Branch (III) - stable steady state solutions for $\theta_Y = 100°$. Width $w=0.375$, distance $d=0.375$, and sharpness factor $p=2$.

*3.1.4 Width effect*

Finally, we examine the effect of varying the stripe width on the wetting behavior of corrugated solid surfaces (see Fig. 8). Here, an increase of the stripe width lowers the critical value of $\theta_Y$ for a "(I) to (III)" branch transition (75° $w$=0.45), as opposed to $\theta_Y$= 84.2° for stripes of width of $w$=0.375. Similarly, the onset of the "(III) to (V)" transition is shifted to lower $\theta_Y$ values, however the shift of the critical $\theta_Y$ value is only of 5° when the width of each individual stripe is increased by 20%.

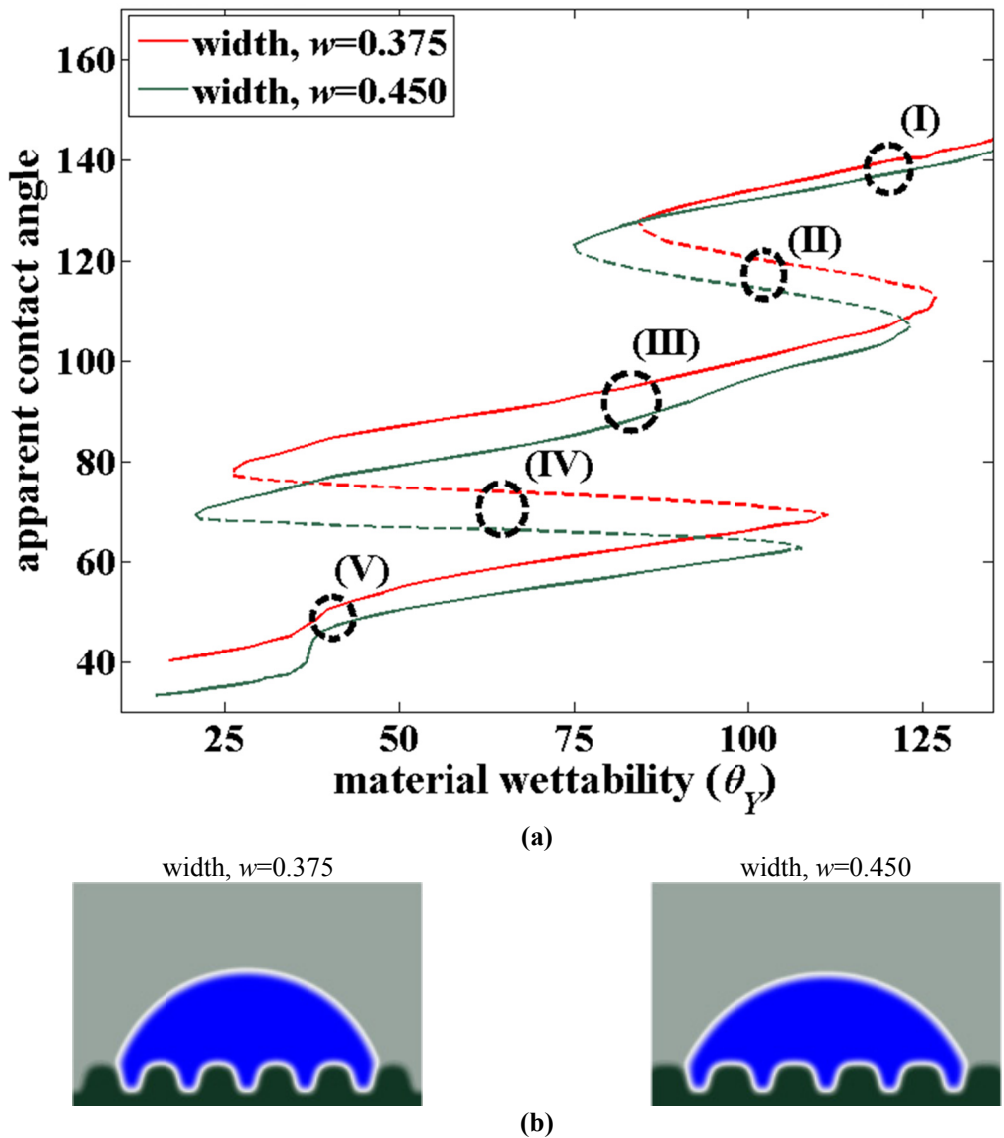

**Figure 8. (a)** Effect of stripe width on the apparent wettability of a corrugated solid surface. **(b)** Branch (V) - collapsed ("Wenzel") steady state solutions for surfaces with $\theta_Y$ =100°. Distance $d$=0.375, height $h$=0.375, and sharpness factor $p$=2.

*3.2 Energy barrier computations*

The presented bifurcation analysis is an essential step towards the design of surfaces with fully controllable wetting properties, since it enables the computation of the energy barriers separating distinct (meta)stable wetting states. In Figs. 9-12, we show the effect of different geometric properties on the energy barriers required to induce wetting ("(I) to (III)", "(III) to (V)" branches) and de-wetting transitions ("(V) to (III)" and "(III) to (I)" branches), between co-existing stable steady states for $\theta_Y = 100°$.

In all studied geometries, we observe low energy barriers for the wetting "(I) to (III)" transitions, suggesting that the metastability of "Cassie"-like states can be characterized as "weak", thus small perturbations can induce such wetting transitions. The energy barrier for wetting "(III) to (V)" transitions is significantly higher than the one for "(I) to (III)" transitions; thus, the impregnating wetting states for a hydrophobic material, with $\theta_Y = 100°$ are more robust. Furthermore, the energy barriers for de-wetting transitions are of comparable magnitude for "(V) to (III)" and "(III) to (I)" transitions. However, the "(III) to (I)" energy barrier is higher in all studied cases, suggesting that the Wenzel impregnating state of branch (III) for $\theta_Y = 100°$ is less sensitive to perturbations than it is the fully wetted Wenzel state. This observation holds for corrugated solid surfaces coated with a hydrophobic material of $\theta_Y = 100°$; for more hydrophilic materials the fully collapsed Wenzel state becomes "more stable", compared to the intermediate Wenzel impregnating state.

Since the droplet is cylindrical, the interfacial energy per unit length $L$ (along the *y*-direction in Fig.3) can be normalized by surface tension $\gamma_{LV}$ and the droplet radius $R$ in order to obtain the dimensionless energy barriers, $EB = E_{i \rightarrow j}/(\gamma_{LV} R L)$ ($i,j$ = I, III, V) reported in Figs. 9-12.

*3.2.1 Stripe edge sharpness effect*

The effect of varying the stripe sharpness on the energy barriers required for the realization of wetting and de-wetting transitions for a hydrophobic material ($\theta_Y = 100°$) is illustrated in Fig. 9. One can observe that an increase of sharpness does not alter significantly the energy barriers for wetting transitions. A small decrease of the energy barrier for a "(I) to (III)" transition is observed, while the transition from the middle to lower branch requires approximately the same amount of energy.

For de-wetting transitions, the energy required for a "(V) to (III)" transition remains practically unchanged when using corrugations of different sharpness. However, a significant change is observed for a de-wetting transition from branch (III) to branch (I). In particular, the middle to upper branch wetting state transition requires significantly higher energy for sharp edge stripes, suggesting an enhancement of the impregnating wetting state robustness by increasing the stripe sharpness.

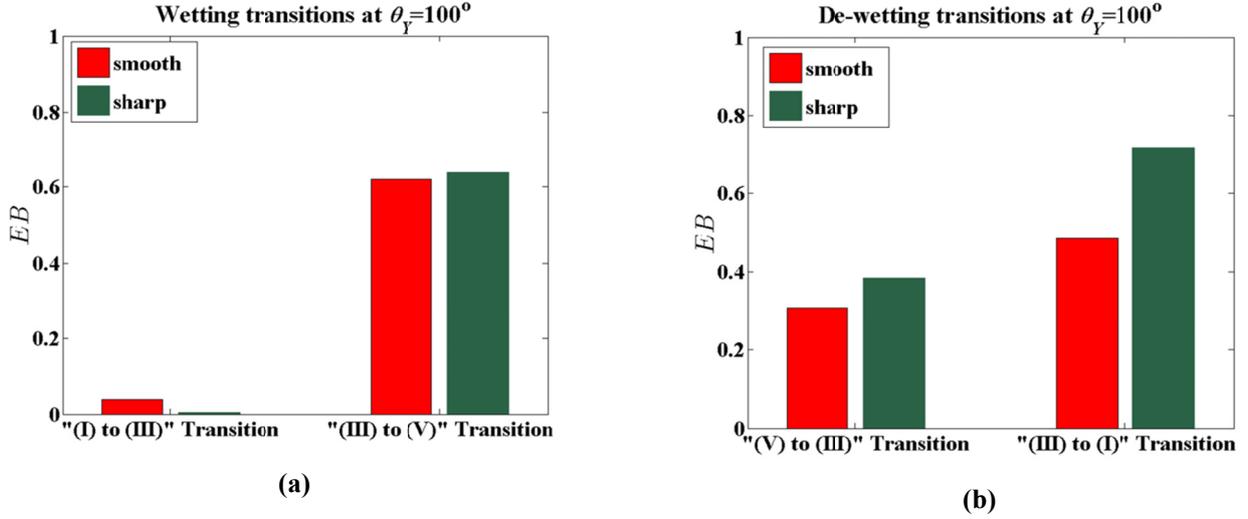

**Figure 9.** Effect of stripe edge sharpness on energy barriers, $EB = E_{i \to j}/(\gamma_{LV} RL)$, for **(a)** wetting and **(b)** de-wetting transitions at $\theta_Y = 100°$. Width $w=0.375$, distance $d=0.375$, and height $h=0.375$.

*3.2.2 Effect of distance between stripes*

The effect of varying the distance on the energy barrier separating co-existing steady states at $\theta_Y = 100°$ is shown in Fig. 10. Increasing the distance between stripes, increases significantly the energy barrier for "(III) to (V)" and "(III) to (I)" de-wetting transitions. The large energy barrier required for a transition originating from branch (III) suggests its robustness to external actuations compared to the co-existing wetting state with a suspended droplet on two (branch (I)) and six (branch (V)) stripes.

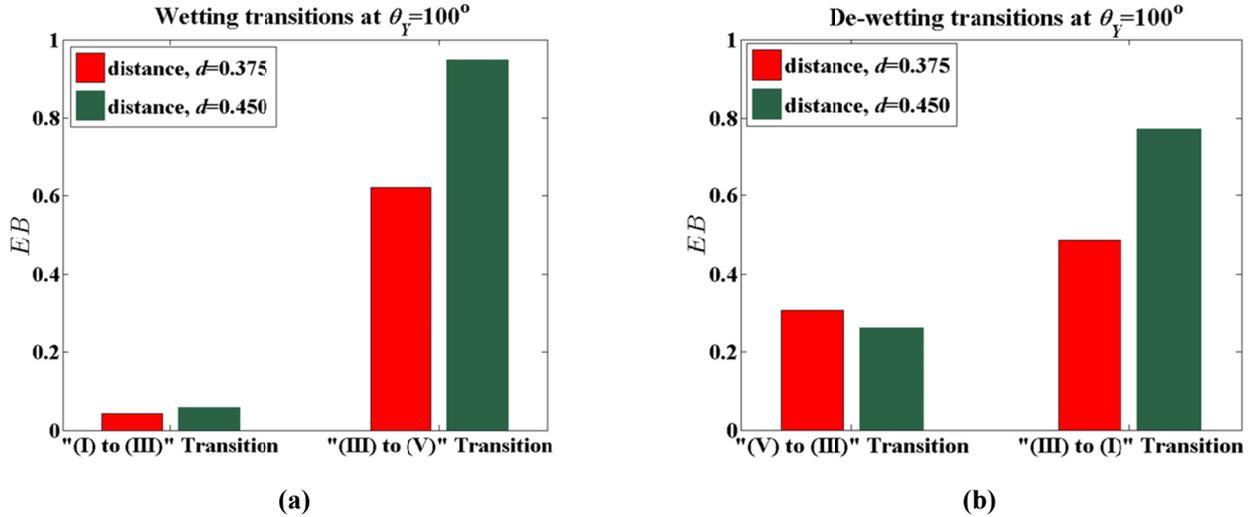

**Figure 10.** Effect of stripe distance on energy barriers, $EB = E_{i \to j}/(\gamma_{LV} RL)$, for transitions between co-existing wetting states at $\theta_Y = 100°$. Width $w=0.375$, height $h=0.375$ and sharpness factor $p=2$.

*3.2.3 Stripe height effect*

An increase of the corrugation depth increases the energy barrier required for a "(III) to (V)" wetting transition for material wettability, $\theta_Y = 100°$ (see Fig. 11). The reverse de-wetting transition from branch (V) to the intermediate (III) branch is also separated by higher energy barrier as we increase the height of the protrusions. Finally, transitions between the upper branch (I) and the stable intermediate branch (III) remain practically unaffected by changing the height of stripes.

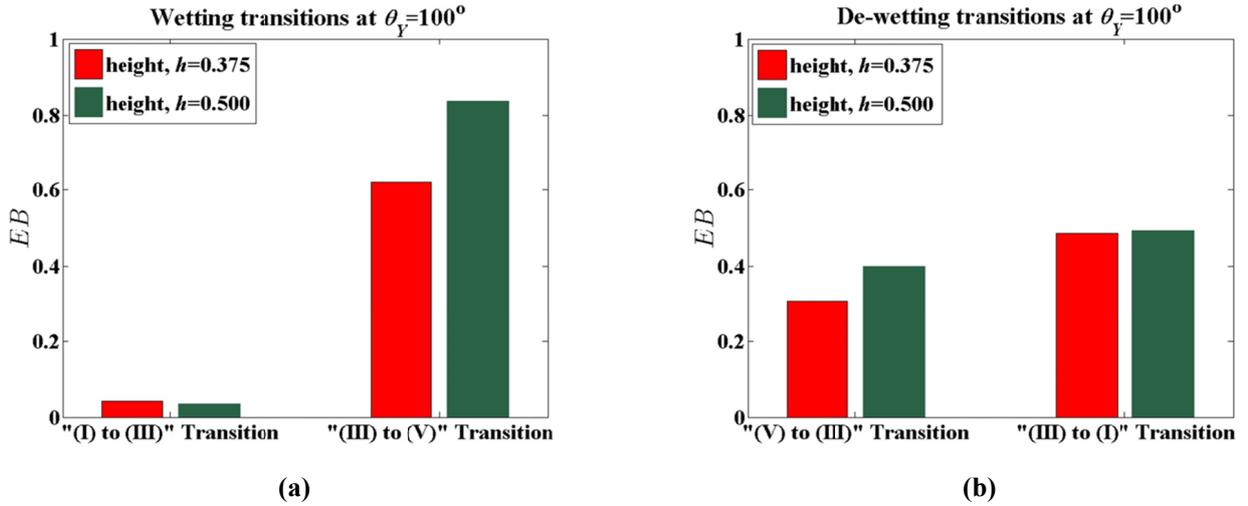

**Figure 11.** Energy barriers, $EB = E_{i \to j}/(\gamma_{LV} RL)$, for **(a)** wetting and **(b)** de-wetting transitions between co-existing steady states at $\theta_Y = 100°$ for stripes of different height. Width $w=0.375$, distance $d=0.375$ and sharpness factor $p=2$.

*3.2.4 Stripe width effect*

Finally, we study the effect of varying the stripe width on the energy barriers separating the co-existing states at $\theta_Y=100°$. The wetting transition from branch (III) to the lower state corresponding to branch (V) is promoted when wide stripes are used (see Fig. 12(a)). On the other hand, de-wetting transitions (see Fig. 12(b)) are favored when stripes of narrower width decorate the solid surface.

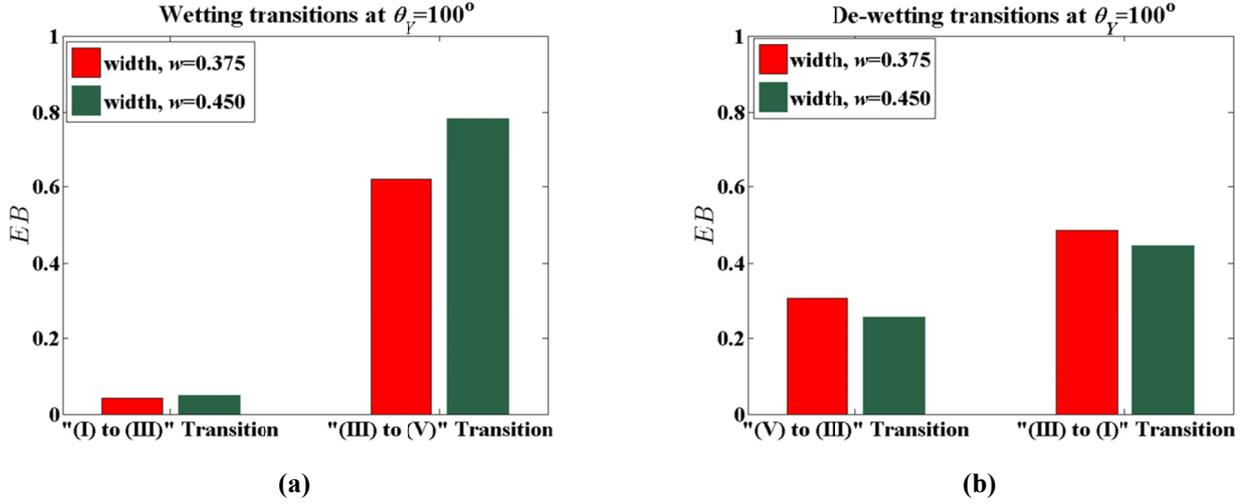

**Figure 12.** Effect of stripe width on energy barriers, $EB = E_{i \to j}/(\gamma_{LV} RL)$, separating co-existing states at $\theta_Y =100^o$. Distance $d$=0.375, height $h$=0.375, and sharpness factor $p$=2.

## 4. Summary and Conclusions

The aim of this work is to quantify the effect of various geometric characteristics of micro-structured solid surfaces on the energy barriers separating co-existing (meta)stable wetting steady states. High energy barriers indicate an increased robustness of the wetting states, whereas low energy barriers are desired when switching between different wetting states is required.

Determination of the energy barriers is made feasible through the computation of all possible, stable and unstable equilibrium solutions that can be admitted on a structured surface. We apply a time-stepper based methodology described in [21], which utilizes short bursts of fine-scale dynamics simulators (e.g. LB simulation [22]) and systems-level analysis including bifurcation and stability analysis. Direct temporal simulations of fine-scale models can only converge to stable steady states after very long executions, however our time-stepper based analysis enables the systematic and efficient exploration of the solution space, computing entire families of both stable and unstable solutions.

The computation of energy barriers for wetting and de-wetting transitions provides a crucial first step for the design of patterned surfaces with specific wetting properties.

The main findings of this analysis are summarized as following:

- Smooth edge stripes facilitate both wetting and de-wetting transitions, through decreasing the corresponding energy barriers
- Increasing the distance between stripes enlarges the range of hysteresis, between suspended and collapsed states.
- Increasing the width of stripes facilitates de-wetting and impedes wetting transitions.
- Wetting and de-wetting transitions are favored when the groove depth is decreased.

It is a matter of available computational power to apply the same methodology to smaller scale structured or unstructured surfaces of increased geometric complexity (e.g., dual scale roughness, honeycomb structures) and engineer the wetting transitions on them by implementing modern minimization algorithms.

Upon minimization of the energy barrier separating the different wetting states, the next step is to design suitable actuations, which can deliver the required energy with minimal losses [39]. Useful information is expected from linear stability analysis. In particular, eigenvectors corresponding to the dominant eigenvalues of the linearized system (eigenvalues of the Jacobian matrix, $\partial \mathit{\Phi}_T(\mathbf{U}^*)/\partial \mathbf{U}^*$) can dictate the optimal directions along which wetting or de-wetting actuations preserve their energy for longer time intervals.

**Acknowledgement**

The authors (MEK & AGP) kindly acknowledge funding from the European Research Council under the Europeans Community's Seventh Framework Programme (FP7/2007-2013) / ERC Grant Agreement no[240710]. CEC was partially founded by the NSF PREM (DMR-0934206). We would also like to thank Prof. I.G. Kevrekidis for valuable discussions.